\documentclass[a4paper,fleqn,usenatbib]{mnras}
\usepackage{amsmath}    % Advanced maths commands
\usepackage[T1]{fontenc}
\usepackage{ae,aecompl}
\usepackage{graphicx}
\usepackage{tabularx}
\usepackage{txfonts}
\usepackage{amssymb}    % Extra maths symbols

\DeclareRobustCommand{\VAN}[3]{#2}
\let\VANthebibliography\thebibliography
\def\thebibliography{\DeclareRobustCommand{\VAN}[3]{##3}\VANthebibliography}

\graphicspath{{./pic/}}

\title[Persistent $\gamma$-ray emission of PKS 1510–089]{
Modelling the persistent low-state $\gamma$-ray emission \\ of the PKS 1510–089 blazar with electromagnetic cascades \\ initiated in hadronuclear interactions}
\author[T.A. Dzhatdoev et al.]{
T. A. Dzhatdoev$^{1,2,3}$\thanks{E-mail: timur1606@gmail.com},
E. V. Khalikov$^{1}$,
V. S. Latypova$^{4}$,
\newauthor
E. I. Podlesnyi$^{1,3,4}$
and I. A. Vaiman$^{1,3,4}$
\\
\\
$^{1}$Federal State Budget Educational Institution of Higher Education, M.V. Lomonosov Moscow State University, \\Skobeltsyn Institute of Nuclear Physics (SINP MSU), 1(2), Leninskie gory, GSP-1, 119991 Moscow, Russia\\
$^{2}$Institute for Cosmic Ray Research, University of Tokyo, \\5-1-5 Kashiwanoha, Kashiwa, Japan\\
$^{3}$Institute for Nuclear Research of the Russian Academy of Sciences, \\7a, 60th October Anniversary Prospect, 117312 Moscow, Russia\\
$^{4}$Federal State Budget Educational Institution of Higher Education, M.V. Lomonosov Moscow State University, \\Department of Physics, 1(2), Leninskie gory, GSP-1, 119991 Moscow, Russia\\
}

\date{Accepted XXX. Received YYY; in original form ZZZ}

\pubyear{2021}

\begin{document}
\label{firstpage}
\pagerange{\pageref{firstpage}--\pageref{lastpage}}
\maketitle
\begin{abstract}
Blazars may accelerate protons and/or nuclei as well as electrons. The hadronic component of accelerated particles in blazars may constitute the bulk of their high-energy budget; nevertheless, this component is elusive due to a high value of the energy threshold of proton interaction with photon fields inside the source. However, broad line regions (BLRs) of some flat spectrum radio quasars (\mbox{FSRQs}) may contain a sufficient amount of matter to render primary protons ``visible'' in $\gamma$ rays via hadronuclear interactions. In the present paper we study the persistent $\gamma$-ray emission of the FSRQ PKS 1510–089 in its low state utilizing the publicly-available \textit{Fermi}-LAT data, as well as using the spectrum measured with the MAGIC imaging atmospheric Cherenkov telescopes. We find an indication for an excess of $\gamma$ rays at the energy range $\gtrsim 20$~GeV with respect to a simple baseline log-parabolic intrinsic spectral model. This excess could be explained in a scenario invoking hadronuclear interactions of primary protons on the BLR material with the subsequent development of electromagnetic cascades in photon fields. We present a Monte Carlo calculation of the spectrum of this cascade component, taking as input the BLR photon field spectrum calculated with the \textit{Cloudy} code. To our knowledge, this is the first calculation of electromagnetic cascade spectrum inside a blazar based on a direct calculation of the photon field spectrum with a spectral synthesis code. \end{abstract}
% Select between one and six entries from the list of approved keywords.
% Don't make up new ones.

%%%%%%%%%%%%%%%%% BODY %%%%%%%%%%%%%%%%%%

\section{Introduction}
\label{sec:introduction}

The extragalactic high-energy (HE, $E > 100$ MeV) $\gamma$-ray sky is dominated by blazars \citep{Abdollahi2020,Wakely2008} --- active galactic nuclei with relativistic jets pointing towards the observer. A large fraction (in some models --- almost an entirety) of non-thermal power radiated by blazars could be attributed to leptonic processes inside the so-called ``blobs'' --- clouds of magnetized plasma propagating along the jets. Besides electrons, protons and/or nuclei could also have been accelerated in blobs/jets of blazars. In 2018, the IceCube collaboration reported on the evidence for astrophysical neutrinos from the blazar TXS 0506+056 \citep{Aartsen2018a,Aartsen2018b}, thus indicating that hadronic processes are indeed in operation in blazars.

Remarkably, TXS 0506+056 was classified by \cite{Padovani2019} as a flat spectrum radio quasar (FSRQ). FSRQs are believed to be powerful (Eddington ratio $L/L_{Edd} > 0.01$) blazars which contain the broad line region (BLR) matter in the form of clouds or wind. High-energy protons accelerated in blobs or jets of FSRQs can sometimes interact with this matter, resulting in observable fluxes of $\gamma$ rays and neutrinos (e.g. \citet{Dar1997,Bednarek1997,Beall1999,Araudo2010,delPalacio2019}).

Accelerated protons are likely to be responsible for the bulk of the high-energy budget in blazars. Nevertheless, in the absence of intervening matter, the protons with the energy below 1--10 PeV may be almost ``sterile'', i.e. in many FSRQs such protons may lose only a small fraction of their energy before escaping from the jet. This conclusion follows from the fact that FSRQs were detected in the very-high-energy domain (VHE, $E > 100$ GeV) with imaging atmospheric Cherenkov telescopes (IACTs) (e.g. \citet{Albert2008,Aleksic2011,Abramowski2013}). These observations imply (at least in the framework of geometrically simple, one-zone emission models) a reasonably low intrinsic $\gamma\gamma\rightarrow e^{+}e^{-}$ pair production (PP) optical depth $\tau_{\gamma\gamma} < 2-3$ below 100 GeV. Indeed, the effective $\gamma$-ray PP energy threshold is $E_{thr-\gamma\gamma} = E_{t}\times (E_{b}/1 \: eV)^{-1}$, where $E_{b}$ [eV] is the characteristic energy of the intrinsic photon field, and $E_{t} \sim 1$ TeV (the energies are given in the blazar's host galaxy rest frame). Therefore, $E_{b} < 10$ eV is implied by the VHE observations. On the other hand, the effective proton energy threshold for the photopion process is $\sim4\:PeV\times (E_{b}/10\:eV)^{-1}$ (e.g. \citet{Berezinsky2006}). We should also note that the energy threshold of the Bethe-Heitler process is somewhat lower. However, neutrinos are not produced in this process. These estimates include all kinds of photon fields, for instance those arising in the spine-sheath configuration (e.g. \citet{Ansoldi2018}); thus, we demonstrate that it is hard to expect observable fluxes of VHE $\gamma$ rays and photohadronic neutrinos from the same blazars in geometrically simple models.

Thus, in order to study the TeV-PeV population of protons inside blazars, it is desirable to search for a signature of hadronuclear (proton-proton, $pp$) interactions in these sources. Inelastic $pp$ interactions have the energy threshold of 1.22 GeV in terms of total energy (see e.g. Fig. 11 of \citet{Kelner2006}). The importance of such a search is still more emphasized by the fact that a significant part of the useful astrophysical neutrino signal of IceCube corresponds to the neutrino energy range of 10-300 TeV, translating to the proton energy range of 100 TeV to 3 PeV (e.g. \citet{Kelner2006,Kelner2008}). Such a choice for the neutrino energy range stems from the fact that below 10 TeV a tremendous atmospheric neutrino background is present, while above 300 TeV the number of expected neutrinos in IceCube is relatively small for most of the conceivable discrete sources.

Such a signature of $pp$ interactions could be present in $\gamma$-ray spectra of some FSRQs in the form of a distinct component at very high energies. Primary protons can interact on material embedded into the broad line region; for instance, $pp$ collisions can take place at the distance from the central black hole of $(0.8-0.9) R_{BLR}$, where $R_{BLR}$ is the characteristic radius of the BLR. In this scenario, the ``residual'' optical depth for $\gamma$ rays exiting the BLR region is \mbox{5--10} times smaller than the ``full'' optical depth for $\gamma$ rays produced near the center of the BLR. Therefore, a part of sub-TeV or even TeV $\gamma$~rays produced near the edge of the BLR could escape the BLR; another part of such $\gamma$ rays could be re-generated in electromagnetic cascades developing in the BLR photon field. This scenario for the 2014-2015 neutrino emission episode of TXS 0506+056 was considered by \citet{Dzhatdoev2019}. In fact, it is a modified scenario of \citet{Dar1997}.

We argue that one of the best candidates for the search of the aforementioned spectral signature is the blazar PKS 1510--089. PKS 1510--089 (cosmological redshift $z = 0.361$) was classified as a FSRQ \citep{Tanner1996}. VHE $\gamma$-ray emission of this source was discovered with the H.E.S.S. telescopes \citep{Abramowski2013}. Persistent VHE $\gamma$-ray emission of PKS 1510--089 over the years from 2012 to 2017 was detected with the MAGIC telescopes \citep{Acciari2018}. Such emission could come directly from the jet, while flares could be attributed to blobs. If the jet of PKS 1510--089 accelerates protons up to the energy of at least several TeV, we expect the existence of a sub-TeV $\gamma$-ray component in the observable spectrum. If the maximal energy of these protons is in excess of 100 TeV, neutrino(s) from PKS 1510--089 could be detected with IceCube or similar neutrino telescopes.

In the present paper we study the persistent HE and VHE $\gamma$-ray emission of PKS 1510--089 in its low state for the case of the primary protons. In Section~\ref{sec:Fermi-LAT-data} we describe the \textit{Fermi}-LAT \citep{Atwood2009} space $\gamma$-ray telescope data analysis for PKS 1510--089 performed by us. Using combined \textit{Fermi}-LAT and MAGIC data, in Section~\ref{sec:excess} we report on a tentative $\gtrsim 20$~GeV excess of $\gamma$ rays with respect to a baseline model. This excess could be attributed to the hadronuclear component discussed above. In Section~\ref{sec:models} we present the 2-component model. In Section~\ref{sec:discussion} we discuss the results and put these results into the context of contemporary blazar studies. Finally, we conclude in Section~\ref{sec:conclusions}.

\section{\textit{Fermi}-LAT data analysis}\label{sec:Fermi-LAT-data}
\subsection{Data selection and analysis setup} \label{subsec:selection}
	
The data were taken from the \textit{Fermi}-LAT data server\footnote{\url{https://fermi.gsfc.nasa.gov/cgi-bin/ssc/LAT/LATDataQuery.cgi}} over the time period from the 4$^{\mathrm{th}}$ of August, 2008 to the 6$^{\mathrm{th}}$ of April, 2020. The region of interest (ROI) is a circle with the radius of $10^{\circ}$ and the center at the position of PKS 1510--089 (4FGL J1512.8-0906 in the 4$^{\mathrm{th}}$ \textit{Fermi}-LAT source catalog 4FGL \citep{Abdollahi2020}). We consider the energy range from 100 MeV to 300 GeV, four logarithmic bins per energy decade. For our analysis, we select events of the \texttt{P8R3 SOURCE} class. Using the \emph{fermipy} package \citep{Wood2017} and assuming the \texttt{P8R3\_SOURCE\_V2} instrument response function set\footnote{\url{https://fermi.gsfc.nasa.gov/ssc/data/analysis/documentation/Cicerone/Cicerone_LAT_IRFs/IRF_overview.html}}, we perform a binned likelihood off-plane point source analysis following the standard recommendations for this type of analysis\footnote{\url{https://fermi.gsfc.nasa.gov/ssc/data/analysis/documentation/Cicerone/Cicerone_Data_Exploration/Data_preparation.html}}.

\begin{figure}
    \centering
    \includegraphics[width=0.475\textwidth]{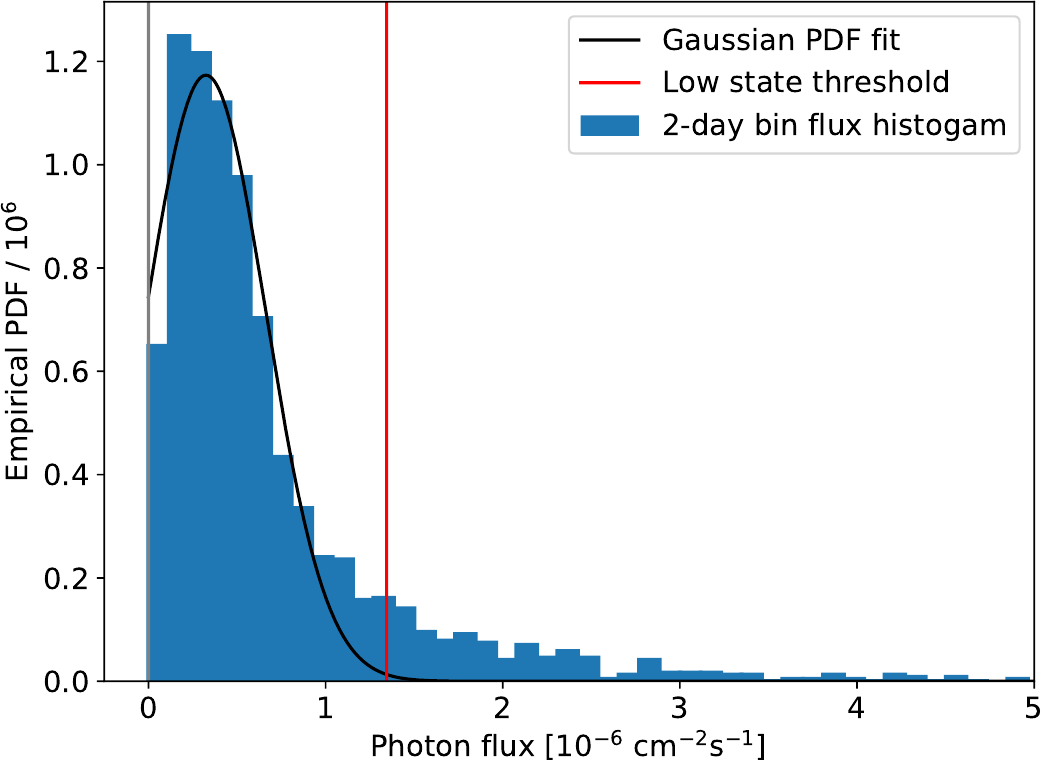}
    \caption{Empirical PDF of photon fluxes in the energy range between 100 MeV and 300 GeV. Black curve represents the Gaussian fit for the PDF, red line represents its $3 \sigma$ quantile that was chosen as a low state photon flux threshold.}
    \label{fig:lowstate_hist}
\end{figure}

\begin{figure*}
\centering
\includegraphics[width=1.0\textwidth]{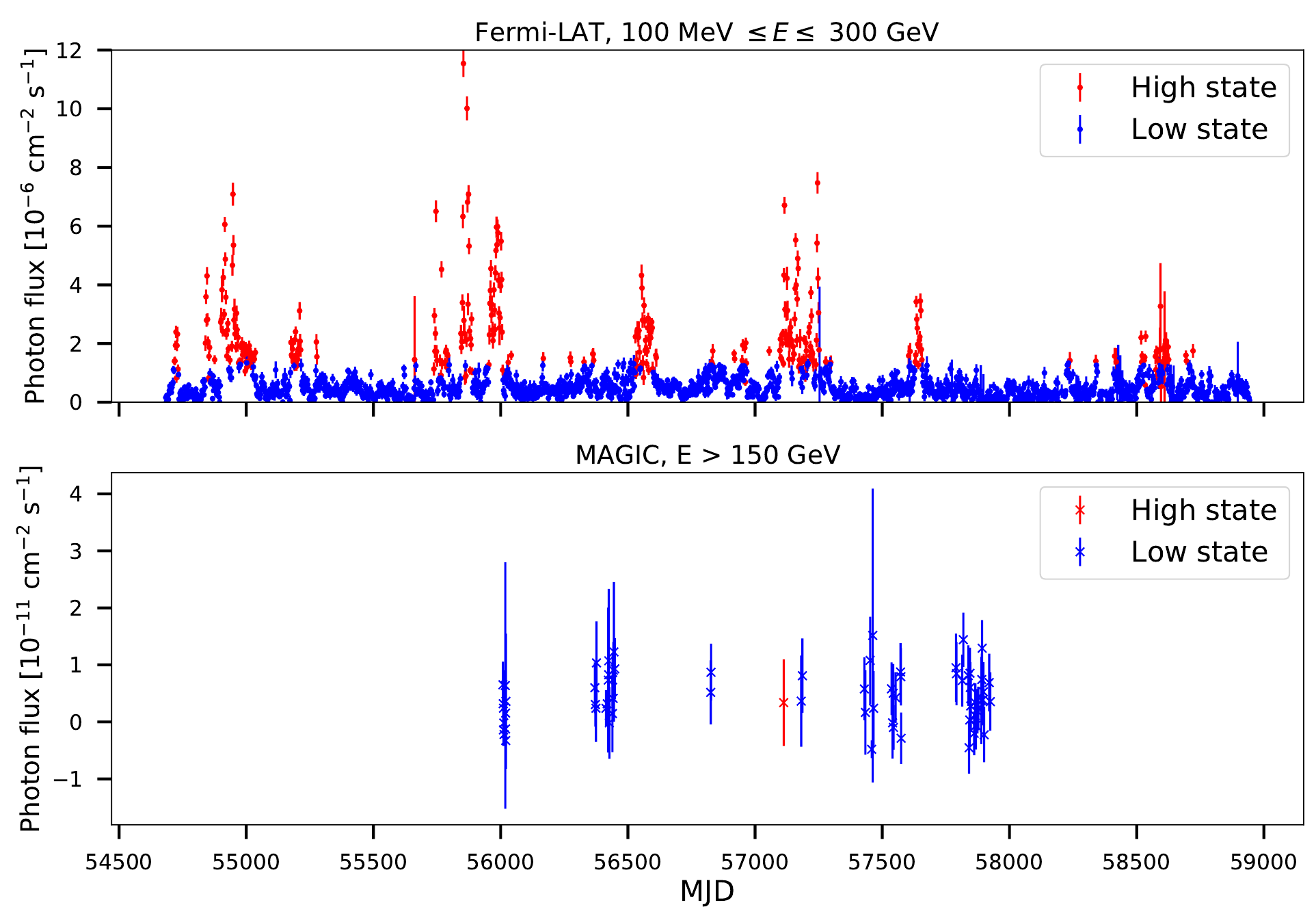}
\caption{Upper panel: the \textit{Fermi}-LAT 2-day bin light curve (our work). Lower panel: the MAGIC low-state light curve \citep{Acciari2018}.}
\label{fig:light_curve}
\end{figure*}

We constructed a model of the observed $\gamma$-ray emission consisting of the following components: 1)~PKS 1510--089, the source of interest; 2)~all other 4FGL sources within a $15^{\circ}$-circle centered at the position of PKS 1510--089; 3)~galactic and isotropic background $\gamma$-ray emissions, represented by the $\texttt{gll\_iem\_v07}$ and $\texttt{iso\_P8R3\_SOURCE\_V2\_v1}$ models, respectively\footnote{\url{https://fermi.gsfc.nasa.gov/ssc/data/access/lat/BackgroundModels.html}}. We modelled the spectrum of PKS 1510--089 with the PowerLaw2\footnote{\url{https://fermi.gsfc.nasa.gov/ssc/data/analysis/scitools/xml_model_defs.html}} function allowing its two parameters (photon flux and spectral index) to vary freely during the model optimization. Photon fluxes of diffuse $\gamma$ rays and 4FGL sources within 7$^{\circ}$ from the ROI center were also left free, as well as the spectral shapes of 4FGL sources within $5^{\circ}$ from the ROI center, while other model parameters, including 4FGL sources' spectral parameters and photon fluxes from sources outside of the radius of 7$^{\circ}$ from the ROI center, were fixed at their catalog values.

After the initial optimization done with the \texttt{fermipy.GTAnalysis.optimize} tool all sources with their test statistic $TS < 16$ or predicted total number of photon counts $N_{\mathrm{pred}} < 10$ were removed from the ROI model. Then, the updated model curves were fitted to the observational data with the \texttt{fermipy.GTAnalysis.fit} tool.

\subsection{PKS 1510--089 low state identification} \label{subsec:light_curve}

In this paper we focus on the persistent $\gamma$-ray emission from PKS 1510--089 in its ``low'' state. Low state, as opposed to high (active or flaring) states, is thought to correspond to relatively stable long-term conditions inside the source. To identify the time intervals corresponding to the low state, we first obtained the \textit{Fermi}-LAT light curve of PKS 1510--089 with a 2-day time binning. 

For the light curve calculation we used the model from Sect. \ref{subsec:selection} additionally fixing the parameters of the galactic and isotropic background $\gamma$-ray emissions to the best-fit values from the analysis of the whole time range, since they are not expected to vary on day timescales. The PKS 1510--089 light curve was obtained with the \texttt{fermipy.GTAnalysis.lightcurve} tool, which additionally fixes the spectral shapes of point sources in the model (except for the source of interest) to their best-fit values from the whole-time-range analysis, but allows to vary their spectrum normalization.

Having obtained the 2-day binned values of the PKS 1510--089 photon flux and spectral index, we made an attempt to employ a 2-dimensional approach to the low state identification using them as independent variables but have not found any significant improvement with respect to flux-only separation. Appendix \ref{two-dim-lowstate-selection} gives more details on this attempt. Hereafter we use photon flux as the only state-related observable.

Then we obtained an empirical probability density function (PDF) of photon fluxes in the same time bins as the ones used for the light curve and fitted the empirical PDF (more precisely, the dependence of the empirical PDF bin content on the value of the photon flux corresponding to the center of the bin) with Gaussian curve. We set the upper $3 \sigma$ quantile of the fitted distribution as a threshold for the low-state photon flux in the energy range between 100~MeV and 300~GeV (see Fig.~\ref{fig:lowstate_hist}). The threshold value is $F_{\mathrm{thr: 100 MeV - 300 GeV}} = 1.34 \times 10^{-6}$~cm$^{-2}$~s$^{-1}$. The median photon flux is $4.7 \times 10^{-7}$~cm$^{-2}$~s$^{-1}$ with the typical photon flux error value $\approx 1.3 \times 10^{-7}$~cm$^{-2}$~s$^{-1}$.

We note that the low state identification procedure performed is rather formal because the low-flux part of the empirical PDF can not be accurately described with a normal distribution since its minimal value is zero. Nevertheless, the Gaussian fit gives a plausible approximation of the empirical PDF to the right from its peak. Moreover, this fit is not affected by the distribution's heavy upper tail that corresponds to the high state.

\begin{figure*}
    \centering
    \includegraphics[width=0.85\textwidth]{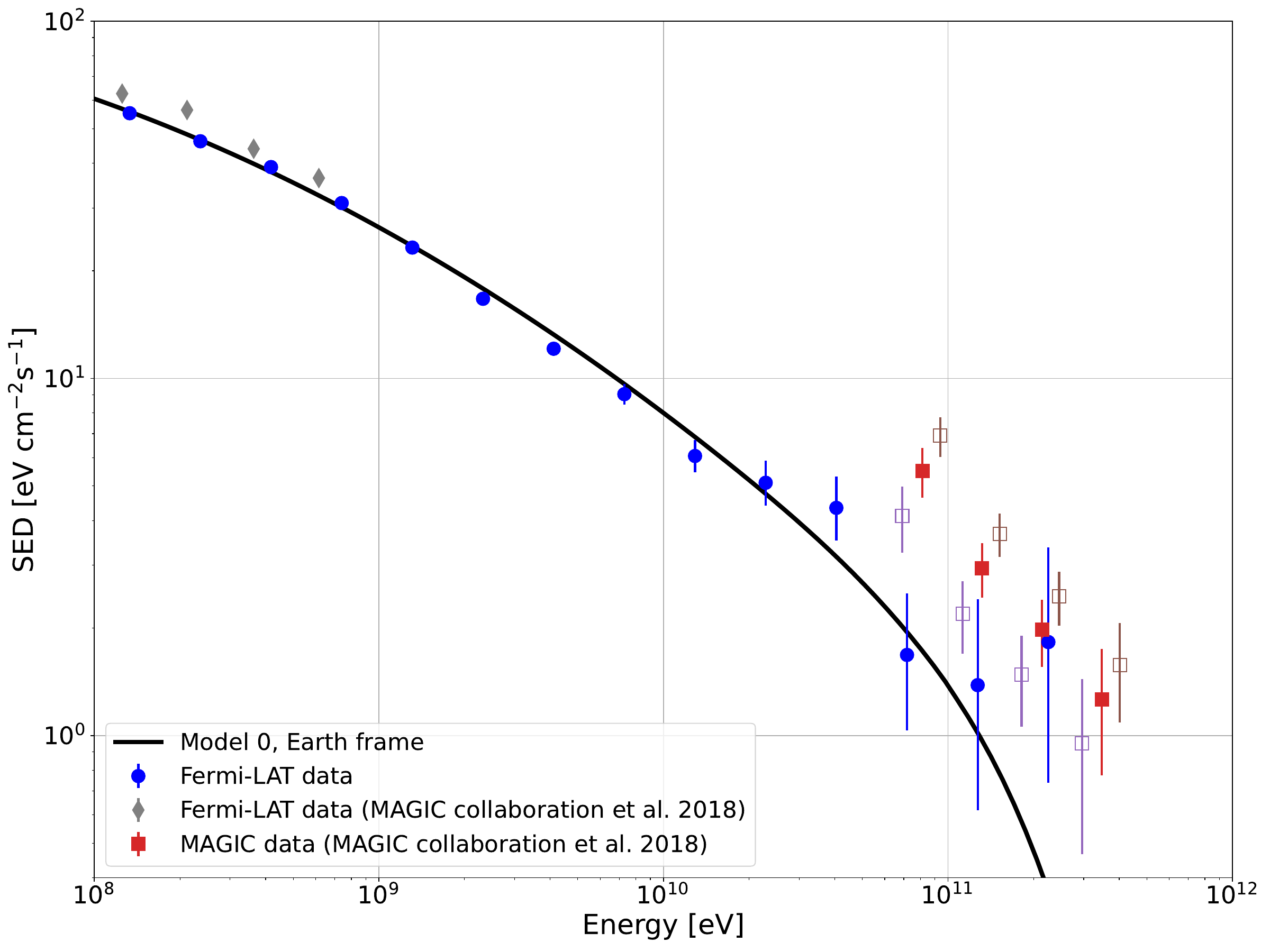}
    \caption{The \textit{Fermi}-LAT (our analysis: blue circles, and the analysis of \citet{Acciari2018}: grey diamonds) and MAGIC (squares, data taken from \citep[][Fig. 7]{Acciari2018}) low-state SEDs of PKS 1510--089. Filled squares represent baseline MAGIC measurements, hollow squares denote the MAGIC systematic uncertainty in energy and SED measurements. Solid curve represents the model 0 with extragalactic absorption taken into account.} \label{fig:SED0}
\end{figure*}

After the threshold of the low state was fixed, we applied a moving average filter with a 6-day (three 2-day bins) window to the light curve in order to exclude transient threshold crossings. To make our low state selection more conservative, we considered the time intervals in which the smoothed light curve value or its actual value exceeds the threshold as belonging to the high state and excluded them from the analysis. The low state intervals were merged into one \texttt{.fits} file and used for spectral analysis. The obtained \textit{Fermi}-LAT light curve with marked low and high states is presented in the upper panel of Fig.~\ref{fig:light_curve}. In the lower panel of Fig.~\ref{fig:light_curve} the MAGIC light curve \citep{Acciari2018} is plotted for the observation nights identified by MAGIC Collaboration as belonging to the low state. We note that one of these items was identified by us as belonging to the high state (shown in red in the lower panel of Fig.~\ref{fig:light_curve}).

\subsection{Spectral energy distribution} \label{subsec:SED}

For the spectral analysis of PKS 1510--089 we used the low-state data for the whole \textit{Fermi}-LAT observation period. We modified our ROI model altering the spectrum shape of the source of interest to that of the log-parabolic function. During the optimization procedure both normalization and spectral shape of PKS 1510--089 were left free. The SED obtained with the \texttt{fermipy.GTAnalysis.sed} method is shown in Fig.~\ref{fig:SED0} as blue circles with statistical error bars. 

The SED of PKS 1510--089 obtained by \citet{Acciari2018} is also presented in Fig.~\ref{fig:SED0}. We note that the \textit{Fermi}-LAT SED obtained by us lies slightly lower than the \textit{Fermi}-LAT SED of \citet{Acciari2018}. This slight discrepancy can be explained with the fact that in our work we apply a more conservative approach to select low-state time intervals as explained in Sect.~\ref{subsec:light_curve}. Taking into consideration both the systematic and statistical uncertainties of the ground-based telescope measurements, \textit{Fermi}-LAT and MAGIC SEDs are in good agreement.

\section{The high-energy excess of \texorpdfstring{$\gamma$}{gamma} rays above 10 GeV} \label{sec:excess}

We fit the measured $\gamma$-ray spectrum of PKS 1510--089 as follows. The intrinsic spectrum (i.e. the spectrum before the intergalactic propagation effects are taken into account) is described with a log-parabolic function:

\begin{equation}
\frac{dN}{dE_{\gamma}} = K_{\gamma} \left(\frac{E_{\gamma}}{E_{\gamma_{0}}} \right)^{-\left(\alpha + \beta \ln(E_{\gamma} / E_{\gamma_{0}}) \right)}, \label{eq:primary_spectrum}
\end{equation}
where $K_{\gamma}$ is the normalization parameter, $E_{\gamma_{0}}$ --- the ``reference energy'', $\alpha$ --- the power-law index, $\beta$ --- the curvature parameter. Furthermore, we account for the effects of $\gamma\gamma$ pair production on the extragalactic background light (EBL) \citep{Nikishov1962,Gould1967} and adiabatic losses (redshift). We assume the EBL model of \citet{Gilmore2012}. The fitting of the observed \textit{Fermi}-LAT and MAGIC SEDs was performed considering $K_{\gamma}$, $\alpha$, and $\beta$ as independent parameters. The value of $E_{\gamma_{0}}$ was fixed at 10 GeV.

We obtained fits for three different MAGIC datasets, namely, baseline measurements (red filled squares in Fig. \ref{fig:SED0}), lower (purple hollow squares), and upper (brown hollow squares) boundary measurements, thus taking into consideration the MAGIC systematic uncertainty. For the first fit we obtained the following values of parameters: $K_{\gamma} = 1.29 \times 10^{-19}$~eV$^{-1}$cm$^{-2}$s$^{-1}$, $\alpha = 2.59$, $\beta = 0.037$. The SED corresponding to this model is denoted in Fig.~\ref{fig:SED0} as model~0. The other two fits obtained almost coincide with the first one and are not shown.

Quality of the obtained fit is rather poor: $\chi^2 / \mathrm{n.d.o.f.} = 5.64$ for the baseline MAGIC measurements, $\chi^2 / \mathrm{n.d.o.f.} = 3.01$ for the lower boundary of MAGIC measurements, and $\chi^2 / \mathrm{n.d.o.f.} = 9.36$ for the upper boundary of MAGIC measurements, where $\mathrm{n.d.o.f.} = 18 - 3 = 15$ is the number of degrees of freedom. The empirical significance of the model 0 rejection is $\approx 6.8 \sigma$ for the baseline MAGIC measurements, $\approx 4.0 \sigma$ for the lower boundary of MAGIC measurements, and $> 8 \sigma$ for the upper boundary of MAGIC measurements.

\begin{figure*}
\centering
    \includegraphics[width=1.0\textwidth]{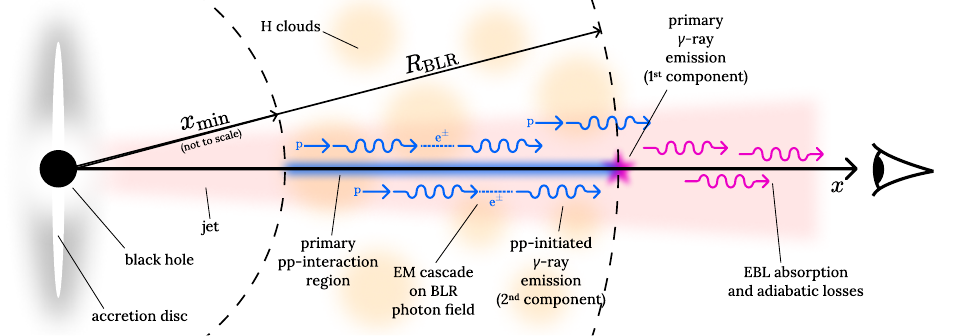}
    \caption{A sketch of the geometry of the FSRQ PKS 1510--089 adopted in our model (not to scale). See text for details.} \label{fig:geometry}
\end{figure*}

The high-energy part of the SED ($\gtrsim 20$~GeV) reveals some excess of $\gamma$~rays above the log-parabolic spectral fit presented in Fig. 3 as black curve. In the next Section we consider a particular model of this excess invoking hadronuclear interactions of primary protons with subsequent development of electromagnetic cascades in photon field.

\section{Hadronuclear interpretation of the $\gtrsim 20$~GeV excess in the low state of {PKS~1510--089}} \label{sec:models}

\subsection{Geometry of the flat-spectrum radio quasar {PKS~1510--089}} \label{subsec:geometry}

A simplified scheme of the geometry of the blazar PKS 1510--089 is shown in Fig.~\ref{fig:geometry}. The source is powered by a central engine represented by a supermassive black hole and an accretion disk. The BLR contains some matter that is shown in Fig.~\ref{fig:geometry} as pale orange circles. This material represents a target for hadronuclear interactions of primary protons. In addition, the BLR material reprocesses radiation from the accretion disk forming the BLR photon field. The outer boundary of the BLR has the radius $R_{\mathrm{BLR}}$. The inner boundary of the BLR material distribution is denoted as $x_{\mathrm{min}}$.

Primary protons are accelerated up to the maximal energy of $\sim 1$ PeV in the jet of the blazar; $pp$ interactions result in the production of $\gamma$ rays at the distance $\leq R_{\mathrm{BLR}}$ from the supermassive black hole. Moreover, electrons and positrons (hereafter collectively called ``electrons'' for simplicity) with energies similar to those of the $\gamma$ rays are produced. In what follows we neglect these electrons and positrons. The effect of this assumption is discussed in Section~\ref{sec:discussion}.

$\gamma$ rays produced in $pp$ interactions initiate electromagnetic cascades on the BLR photon field. These hadronuclear $\gamma$ rays are considered as primary particles for electromagnetic cascades, but they are secondary particles with respect to the protons producing them. Observable $\gamma$ rays from electromagnetic cascades may explain the high-energy excess reported in the previous Section. Finally, the low-energy (log-parabolic) component may be produced at the edge of the BLR or outside of the BLR, as assumed in some contemporary FSRQ $\gamma$-ray emission models (see e.g. \citet{Meyer2019}).

\subsection{The spectrum of the BLR photon field} \label{subsec:blr-spectrum}

The spectrum of the photon field inside the broad line region was calculated with the \textit{Cloudy} code\footnote{\url{www.nublado.org}}, version c17.00 \citep{Ferland2017}. The following values of input parameters were assumed: the bolometric luminosity of the accretion disk $L_{\mathrm{D}} = 10^{46}$~erg/s, the BLR radius $R_{\mathrm{BLR}} = 0.036$~pc, the BLR material covering factor $f = 0.1$, the hydrogen density $h_{\mathrm{den}} = 10^{10}$~cm$^{-3}$, and the column density $N_c = 10^{23}$~cm$^{-2}$. This value of $N_c$ is typical in BLR models (see, e.g., \citet{Kwan1981}). \citet{Miniutti2014} derive $N_c = 10^{23}-10^{24}$~cm$^{-2}$ for the Seyfert 1 galaxy ESO 323–G77 from X-ray observations. The spectrum of the accretion disk was assumed to follow the model ``AGN'' implemented in \textit{Cloudy} with the characteristic temperature parameter amounting to $10^{5}$ K. Other parameters were set to their default values. 

The BLR radius for PKS 1510--089 was measured by \citet{Rakshit2020} with the reverberation mapping technique. The value of $L_{\mathrm{D}}$ assumed by us is between the estimates of \citet{Abdo2010} ($3 \times 10^{45}$~erg/s) and \citet{Rakshit2020} (2.15$\times10^{46}$~erg/s). The values of other parameters are typical for FSRQ studies \citep[see, e.g.,][]{Tavecchio2008}.

After running the \textit{Cloudy} simulation, we obtained the spectral density for the photon field reflected into the BLR $n(E')$ (in units of eV$^{-1}$ cm$^{-3}$), which was used as a background photon field for high-energy $\gamma$ rays. For simplicity, we assume that this photon field is homogeneous and isotropic. Detailed calculations of \citet{Abolmasov2016} show that this is indeed a reasonable first-order approximation.

\subsection{2-component $\gamma$-ray emission model for the low state of {PKS 1510--089}} \label{subsec:2-component}

\begin{figure*}
    \centering
    \includegraphics[width=0.85\textwidth]{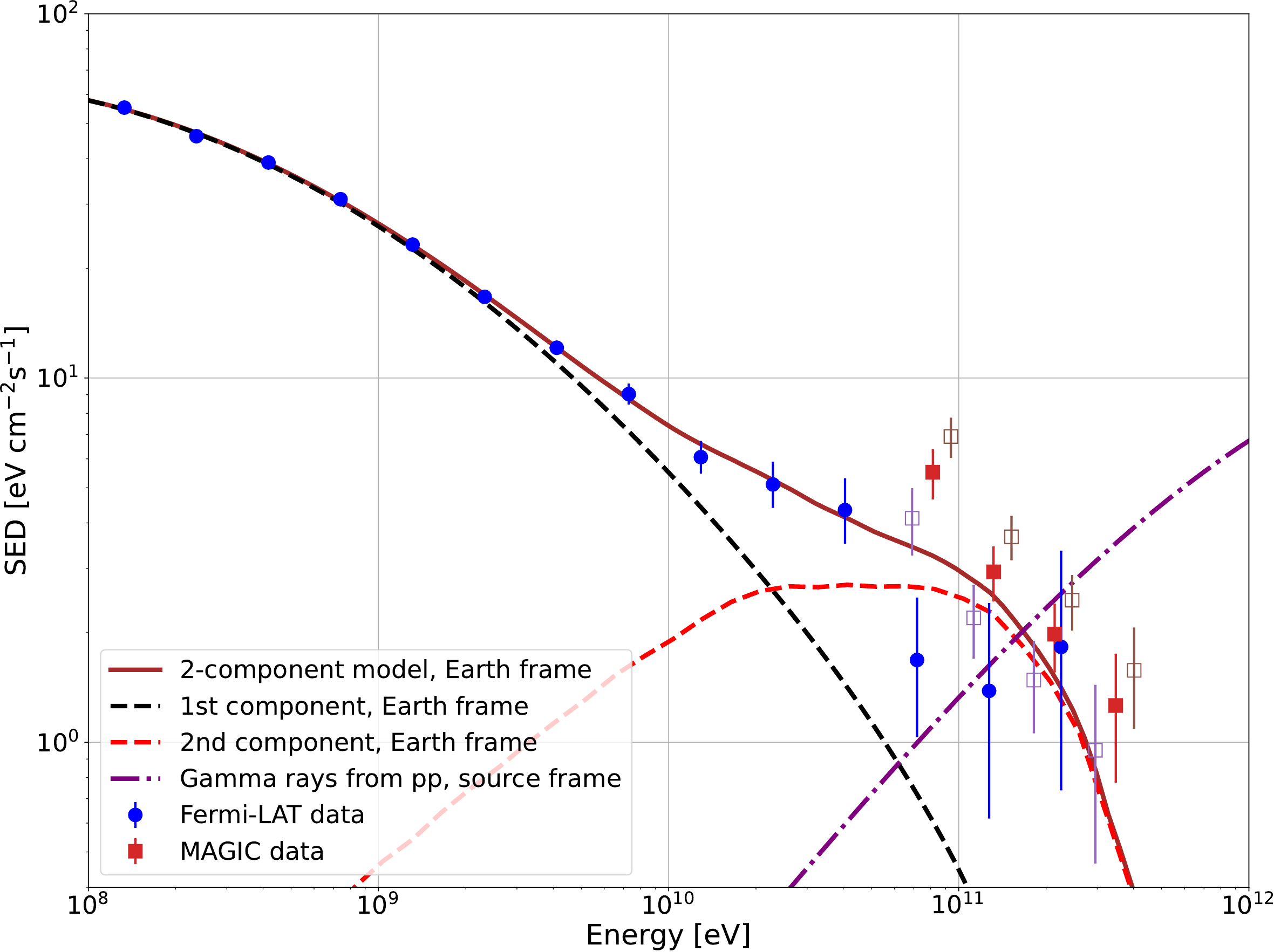}
    \caption{Spectral energy distribution of PKS 1510--089. Observational data points are the same as in Fig. \ref{fig:SED0}. Solid brown curve denotes the 2-component model SED. Separate components of the model are also shown as dashed curves (see plot legend). Dot-dashed purple curve represents the SED of primary (for electromagnetic cascades) $\gamma$ rays produced in proton-proton ($pp$) collisions in the source frame; this curve peaks at the energy of $\approx 10^{13}$~eV.}
    \label{fig:SED2}
\end{figure*}
\begin{figure}
	\centering
	\includegraphics[width=0.490\textwidth]{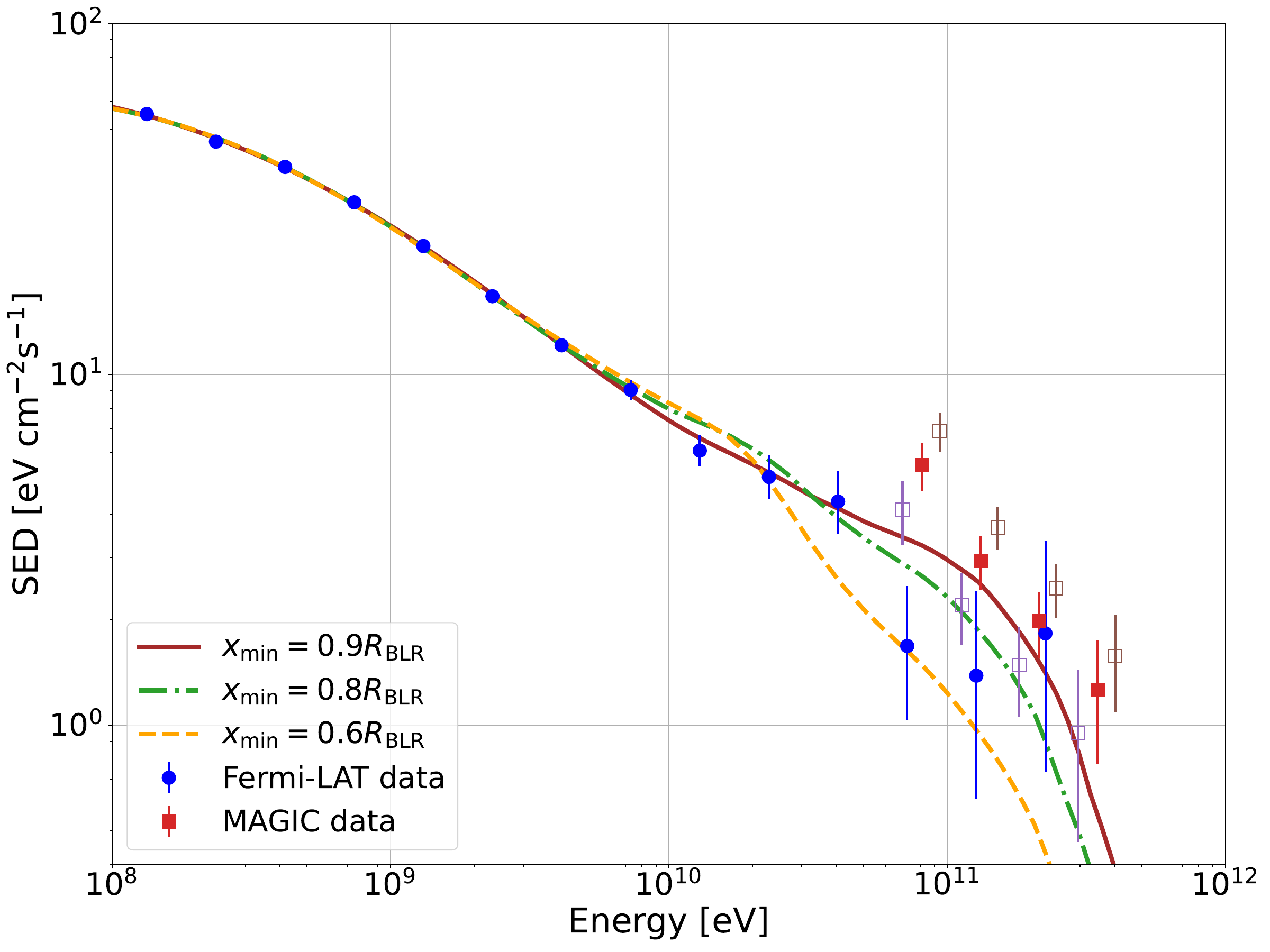}
	\caption{Spectral energy distribution of PKS 1510–089 fitted with the 2-component model for three different values of $x_{\mathrm{min}}$ (see legend).}
	\label{fig:60-80-90}
\end{figure}

Here we describe our calculation of the observable $\gamma$-ray spectrum with an additional, 2nd component from $pp$ interactions. We assume a power-law spectrum of the primary protons: 
\begin{equation}
\frac{dN_p}{dE_p} = K_p \left(\frac{E_p}{E_{p_0}}\right)^{-\gamma_p},
\end{equation}
where proton energy $E_p$ lies in the range between $E_{p_{\mathrm{min}}} = 10^{13}$~eV and $E_{p_{\mathrm{max}}} = 10^{15}$~eV; $\gamma_p = 1.8$ and $E_{p_0} = 10^{10}$~eV.

In what follows we use approximations of \citet{Kelner2006} for the proton-proton inelastic cross section $\sigma_{pp} (E_{p})$ and the secondary $\gamma$-ray spectrum $dN_{\gamma_{s}}/dE_{\gamma_{s}}$. For the matter column density $N_c = 10^{23}$~cm$^{-2}$, the optical depth for $pp$ interactions is 
\begin{equation}
    \tau_{pp} (E_{p}) = N_c \sigma_{pp} (E_{p}).
\end{equation}
The spectrum of interacted primary protons, which produce secondary $\gamma$ rays, is calculated as
\begin{equation}
    \left( \frac{dN_p}{dE_p} \right)_{\mathrm{int}} = \frac{dN_p}{dE_p} \left(1 - \exp{[-\tau_{pp}(E_p)]}\right).
\end{equation}
Furthermore, we assume that the BLR matter is uniformly distributed in the range $x_{\mathrm{min}} \leq x_{p} \leq R_{\mathrm{BLR}}$, where $x_{\mathrm{min}}~=~0.9 R_{\mathrm{BLR}}$ (we have also tried $x_{\mathrm{min}}~=~0.6R_{\mathrm{BLR}}$ and $x_{\mathrm{min}}~=~0.8R_{\mathrm{BLR}}$, but these options led to poorer agreement of the obtained model SED with the data). The $\gamma$-ray production site for the low-energy component with the log-parabolic spectrum is located at $R_{\mathrm{BLR}}$ both for the model 0 and the 2-component model, which results in the absence of the $\gamma\gamma$-absorption of $\gamma$ rays of the 1st (log-parabolic) component on the BLR photon field.

Most of $\gamma$ rays produced in hadronuclear interactions of the primary protons with the BLR material (with the optical depth of this process being $\approx 5 \times 10^{-3}$ for $N_c = 10^{23}$~cm$^{-2}$) escape into the BLR photon field, propagate further along the jet (see blue waves in Fig. \ref{fig:geometry}), and initiate electromagnetic cascades on the BLR photon field. A small part of $\gamma$ rays produce electron-positron pairs on the BLR material; in what follows, we do not account these interactions. $\gamma$ rays that have escaped from the BLR material typically produce $e^{+}e^{-}$ pairs in the ordered magnetic field of the jet; in this case, additional deflection of the secondary (cascade) electrons could be neglected. A detailed investigation of the magnetic field structure in the jet is beyond the scope of the present paper.

The development of electromagnetic cascades proceeds as follows. Most of the hadronuclear $\gamma$ rays produce electron-positron pairs on the BLR photon field with the energy density distribution $n(E')$. For these $\gamma$ rays the optical depth of the PP process on the BLR photon field exceeds unity. Electrons produced in the PP process are shown as blue dotted lines in Fig. \ref{fig:geometry}. These electrons are, in turn, subject to the inverse Compton (IC) process \citep[see, e.g.,][]{Blumenthal1970a} resulting in the transfer of energy from them to the photons of the BLR photon field, thus producing the next-generation $\gamma$ rays of the cascade (subsequent blue waves in Fig. \ref{fig:geometry}). The cascade may comprise one or more generations.
	
To calculate the spectrum of electromagnetic cascades we developed a Monte Carlo code, for the most part of this work following the prescriptions of \citet{Kachelriess2012} for the total and differential cross sections of the PP and IC processes. We neglect the synchrotron energy losses for cascade electrons. We note that the difference of our algorithm from the one presented by \citet{Kachelriess2012} is relatively small.

Using this code, we calculated the spectra of observable $\gamma$ rays for $N_0 = 30 000$ cascades with the primary $\gamma$-ray energy distributed according to the power-law spectrum with the spectral index $-1$ and energies varying in the range from $10^{9}$~eV to $10^{15}$~eV. The energies of all primary and cascade $\gamma$ rays which reach the edge of the BLR (when $x = R_{\mathrm{BLR}}$) and become observable are recorded. For all observable cascade $\gamma$ rays information about the energy of the primary $\gamma$ ray, which initiated the development of a particular electromagnetic cascade, is collected. This allows us to reweigh the resulting spectrum of observable $\gamma$ rays according to the primary spectrum $dN_{\gamma_{s}}/dE_{\gamma_{s}}$ of $\gamma$ rays produced in $pp$ interactions.

The low-energy component has three free parameters $(K_{\gamma}, \alpha, \beta)$. Concerning the high-energy component, the values of $(E_{p_{\mathrm{min}}}, E_{p_{\mathrm{max}}}, \gamma_p)$ were fixed above. The only free parameter remaining is the normalization of the proton spectrum $K_p$. We performed the fitting of the 2-component model to the observational data and obtained the following best-fit values of the parameters: $K_{\gamma} = 9.64 \times 10^{-20}$~eV$^{-1}$cm$^{-2}$s$^{-1}$, $\alpha = 2.80$, $\beta = 0.072$, $K_p = 2.27 \times 10^{-17}$~eV$^{-1}$cm$^{-2}$s$^{-1}$.

The observable $\gamma$-ray component from cascades (the 2nd component) is shown in Fig.~\ref{fig:SED2} as a dashed red curve. In this figure we present the lower-energy component (dashed black curve) and the total observable $\gamma$-ray SED (solid brown curve) as well. The latter SED is the sum of the two former components. For all the three presented SEDs the effect of the extragalactic absorption of $\gamma$ rays was taken into account assuming the EBL model of \citet{Gilmore2012}. For the obtained fit with the 2-component model $\chi^2 / $n.d.o.f.~$= 2.0$ for the baseline MAGIC measurements, $\chi^2 / $n.d.o.f.~$= 1.0$ for the lower boundary of MAGIC measurements, and  $\chi^2 / $n.d.o.f.~$= 4.1$ for the upper boundary of MAGIC measurements, where n.d.o.f. $= 18 - 4 = 14$. Taking into consideration the systematic uncertainties of the MAGIC measurements, we conclude that the 2-component model describes observational data reasonably well over the whole energy range available.

A comparison of the 2-component model SED for the case of $x_{\mathrm{min}} = 0.9 R_{\mathrm{BLR}}$ with model SEDs for the cases of $x_{\mathrm{min}} = 0.8R_{\mathrm{BLR}}$ and $x_{\mathrm{min}} = 0.6 R_{\mathrm{BLR}}$ is presented in Fig.~\ref{fig:60-80-90}. The value of $0.9R_{\mathrm{BLR}}$ yields the best agreement with the data.

The apparent isotropic luminosity of the 2nd (high-energy) $\gamma$-ray component in the host galaxy rest frame is $L_{\gamma-2-\mathrm{iso}} = 3.5\times10^{46}$~erg/s. For the $pp$ interaction optical depth $\tau_{pp}~=~5~\times~10^{-3}$ and the fraction of the primary proton energy channeled into secondary $\gamma$ rays $f_{\gamma}~=~0.2$ (e.g. \citet{Knapp1996}), the apparent isotropic luminosity of the protons is $L_{p-\mathrm{iso}}~=~L_{\gamma-2-\mathrm{iso}}/(\tau_{pp}f_{\gamma}) = 3.5\times10^{49}$~erg/s. Accounting for the anisotropy of the primary protons in the host galaxy rest frame, we get the following one-jet luminosity in the same rest frame:
\begin{equation}
    L_{p} \approx L_{p-\mathrm{iso}} \left(1-\cos(\theta_{\mathrm{jet}})\right) \left(\frac{N_{c}}{10^{23}\mathrm{cm}^{-2}} \right)^{-1} \approx L_{p-\mathrm{iso}} \frac{\theta_{\mathrm{jet}}^{2}}{2} \left(\frac{N_{c}}{10^{23}\mathrm{cm}^{-2}} \right)^{-1},
\end{equation}
where $\theta_{\mathrm{jet}}$ is the jet half-opening angle. According to \citet{Jorstad2017,Weaver2022}, $\theta_{\mathrm{jet}}$ is about 1$^{\circ}$. Therefore, $L_{p}~\approx~5 \times 10^{45}$ erg/s assuming $N_{c} = 10^{23}$ cm$^{-2}$. \citet{Ghisellini2014} find that the kinetic power of relativistic jets from active galactic nuclei is significantly larger than the
luminosity of their accretion disks. Therefore, we conclude that the obtained value of $L_{p}$ is relatively modest.

\section{Discussion} \label{sec:discussion}

In this paper, we have found an indication for an excess of $\gamma$ rays at the energy $E \gtrsim 20$~GeV for the low state of PKS 1510--089. In the previous Section we have described a specific model of this excess involving hadronuclear interactions of multi-TeV protons with the subsequent development of electromagnetic cascades on the BLR photon field. This model is supported with an observation of a particular case of jet-cloud interactions in radio galaxy 3C 84 that was recently reported by \citet{Kino2021}. 

Most of works on blazar $\gamma$-ray emission ignore electromagnetic cascades. This phenomenon was occasionally studied in context of emission models of various types of active galactic nuclei \citep{Blandford1995,Neronov2007,Aharonian2008,Roustazadeh2010,Wendel2021a,Wendel2021b}. However, to the best of our knowledge, the present work is the first paper where the simulation of electromagnetic cascades in the BLR photon field is based on a direct calculation of the photon field spectrum with a spectral synthesis code such as \textit{Cloudy}.

We note that our model could be further elaborated in a number of ways discussed below.
\begin{enumerate}
\item It was assumed that the radius of the proton interaction site is much smaller than the BLR radius. This simplifying assumption could be reconsidered in the future, including $x_{p}$ as another free parameter of the model and allowing for an extended proton interaction site (see, e.g., \citet{Potter2012} for a discussion of blazar models with extended emission zones).
\item The lower-energy component of the 2-component model is parametrized as intrinsic $\gamma$-ray spectrum with the log-parabolic shape. This assumption is justified if $\gamma$ rays are produced via IC scattering of relativistic electrons, which gained energy as a result of the stochastic acceleration \citep{Massaro2003, Tramacere2011}. In this paper we considered the case without the intrinsic $\gamma$-ray absorption of the lower-energy component, corresponding to the case of the emission zone at the border or outside of the BLR. The opposite case is also worth studying.
\item Turbulent magnetic field present near the proton interaction site could deflect the primary protons and electrons produced in $pp$ collisions, broadening the angular distribution of cascade $\gamma$ rays and thus lowering their effective contribution to the observable intensity (given that blazars are believed to be strongly beamed $\gamma$-ray sources). In particular, this deflection effect could be responsible for the effective low-energy cutoff in the proton spectrum characterized by the $E_{p_{\mathrm{min}}}$ parameter. The contribution of cascade $\gamma$ rays from hadronuclear electrons strongly depends on the magnetic field parameters.
\item In this paper we focused on the HE $\gamma$-ray domain ($E > 100$ MeV). Broadband modelling of the spectrum of PKS 1510--089 \citep[see, e.g.,][]{Roy2021} may be of great importance, since it could yield constraints on the strength of the magnetic field inside the jet. In this paper we neglected the synchrotron energy losses for cascade electrons. For the lower-energy (log-parabolic) component this assumption is, indeed, justified, because the $\gamma$-ray luminosity (due to the IC process) is much higher than the radio luminosity (due to synchrotron radiation) (see Fig. 6 of \citet{Acciari2018}). However, a significant part of the energy of cascade electrons responsible for the higher-energy $\gamma$-ray component could be channeled into synchrotron photons. The inclusion of the synchrotron losses could render the model more self-consistent and more precise.
\item In this work we aimed to show that the apparent excess of $\gamma$~rays could be due to $pp$ interactions inside the jet. This is a natural scenario given the growing evidence for IceCube neutrino association with blazars \citep{Aartsen2018a,Aartsen2018b,Plavin2020,Plavin2021} (however, \citet{Zhou2021} have questioned the association of IceCube neutrinos with radio-bright AGN). We note that at least a part of astrophysical neutrinos detected with IceCube could have been generated in $pp$ collisions in FSRQs. However, photohadronic interactions of $\ge10$~PeV protons could also contribute to the $\gamma$-ray excess.
\item The IceCube collaboration reported on the detection of 13$\pm$5 muon-neutrino events above the expected background during the 5-month low-state emission period of the blazar TXS 0506+056 in 2014-2015 \citep{Aartsen2018b}. The low-state neutrino emission from PKS 1510--089 could also be potentially detected with IceCube. The expected number of events from the latter source depends on many factors, such as the energy spectrum of the primary protons and the angular distributions of the neutrino and $\gamma$-ray components, which could be somewhat different. Neglecting the difference between these angular distributions and assuming the effective area for muon neutrinos according to \citet{Aartsen2017}, we estimate the number of IceCube events during the MAGIC observation period (six years) to be between 4 and 15 events. The former value corresponds to the cutoff in the primary proton spectrum $E_{p_{\mathrm{max}}} = 200$ TeV, the latter --- $E_{p_{\mathrm{max}}} = 1$ PeV. The typical background during the same observation period is about 30 events (see Fig. 2 of \citet{Aartsen2018b}).
\end{enumerate}

We note that the spectral shape of the cascade component depends weakly on the primary proton spectrum parameters $(\gamma_p$, $E_{p_{\mathrm{min}}}$, $E_{p_{\mathrm{max}}})$. This is due to the so-called ``cascade universality'' property \citep{Berezinsky2016}, namely, the observable spectrum of cascade $\gamma$ rays is virtually independent of the parent $\gamma$-ray spectrum and the type of the parent particle ($\gamma$ ray/electron). As was shown above, the contribution of $\gamma$ rays produced in cascades initiated by hadronuclear electrons is uncertain. However, due to the cascade universality, the shape of this additional $\gamma$-ray component is nearly the same as the one for the case of cascades from the primary $\gamma$-rays.

In this paper we performed a \textit{Fermi}-LAT data analysis of the low-state data for the particular FSRQ, namely, the blazar PKS 1510--089. A similar high-energy excess search could be performed for other FSRQs in order to establish whether the reported $\gamma$-ray excess at energies $E \gtrsim 20$~GeV is common for other FSRQs in their low states.

\section{Conclusions} \label{sec:conclusions}
In this work we studied persistent $\gamma$-ray emission of the blazar PKS 1510--089. We performed a \textit{Fermi}-LAT data analysis identifying the low-state periods of PKS 1510--089 $\gamma$-ray emission and combining \textit{Fermi}-LAT and MAGIC observations to obtain the SED of the blazar in the energy range from $100$~MeV to $\approx 500$~GeV. We found an indication of the $\gamma$-ray excess at energies $E \gtrsim 20$~GeV (the conservative estimate of the statistical significance gives $\approx 4.0 \sigma$). The observed SED could be reasonably well fitted in the framework of the 2-component model with a lower-energy component well described with the log-parabolic intrinsic spectrum, and the higher-energy component resulting from $pp$ collisions of 10--1000~TeV protons interacting with the BLR material followed by the development of electromagnetic cascades in the BLR photon field.

\section*{Acknowledgements}
The reported study was funded by RFBR, Russia, project number 20-32-70169. E.I.P. thanks the Foundation for the Advancement of Theoretical Physics and Mathematics ``BASIS'' (Contract No. 20-2-10-7-1) and the Non-proﬁt Foundation for the Advancement of Science and Education ``INTELLECT'' for the student scholarships.
%%%%%%%%%%%%%%%%%%%%%%%%%%%%%%%%%%%%%%%%%%%%%%%%%%

\section*{Data Availability}

The data underlying this article will be shared on reasonable request
to the corresponding author.
 
%The inclusion of a Data Availability Statement is a requirement for articles published in MNRAS. Data Availability Statements provide a standardised format for readers to understand the availability of data underlying the research results described in the article. The statement may refer to original data generated in the course of the study or to third-party data analysed in the article. The statement should describe and provide means of access, where possible, by linking to the data or providing the required accession numbers for the relevant databases or DOIs.

%%%%%%%%%%%%%%%%%%%% REFERENCES %%%%%%%%%%%%%%%%%%

% The best way to enter references is to use BibTeX:

\bibliographystyle{mnras}
\bibliography{BLR_cascades} % if your bibtex file is called example.bib

% Alternatively you could enter them by hand, like this:
% This method is tedious and prone to error if you have lots of references
%\begin{thebibliography}{99}
%\bibitem[\protect\citeauthoryear{Author}{2012}]{Author2012}
%Author A.~N., 2013, Journal of Improbable Astronomy, 1, 1
%\bibitem[\protect\citeauthoryear{Others}{2013}]{Others2013}
%Others S., 2012, Journal of Interesting Stuff, 17, 198
%\end{thebibliography}

%%%%%%%%%%%%%%%%%%%%%%%%%%%%%%%%%%%%%%%%%%%%%%%%%%

%%%%%%%%%%%%%%%%% APPENDICES %%%%%%%%%%%%%%%%%%%%%

\appendix

\section{Low state identification using 2D classification with spectral index and photon flux} \label{two-dim-lowstate-selection}

\begin{figure}
	\centering
	\includegraphics[width=0.475\textwidth]{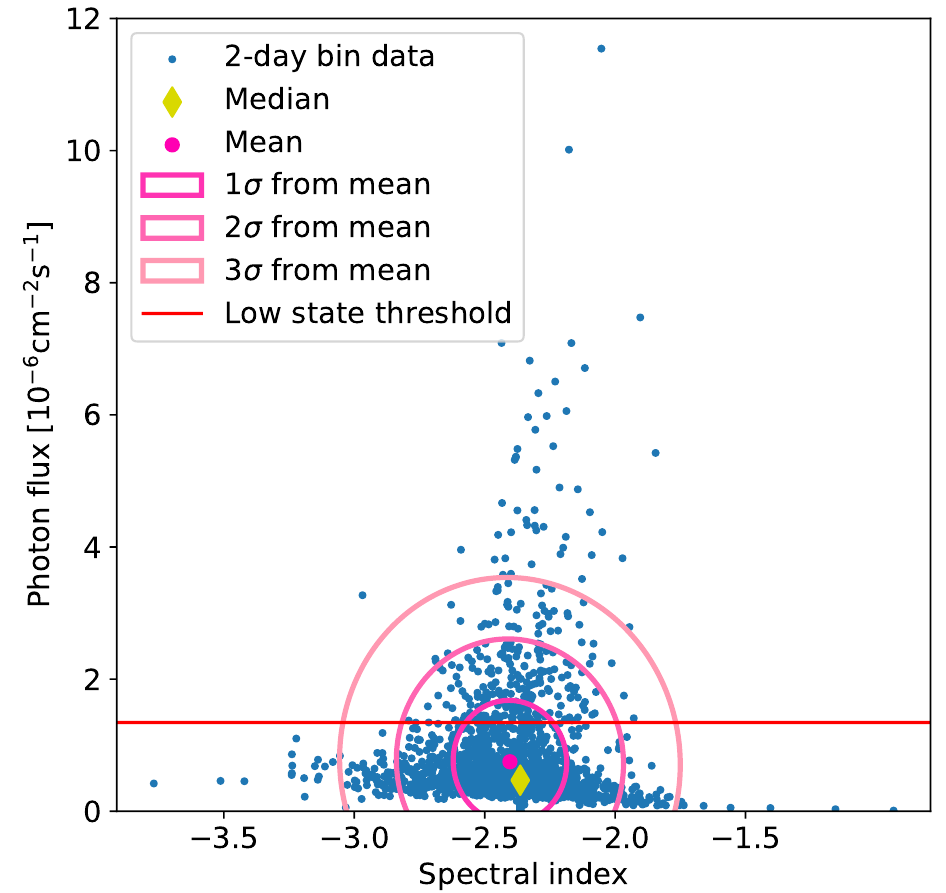}
	\caption{Scatter plot of independent spectral parameters. Each dot represents values of the spectral index and the photon flux (in the energy range between 100 MeV and 300 GeV) corresponding to two days of observations.}
	\label{fig:2d_lowstate}
\end{figure}

In addition to the low-state identification procedure described in Section \ref{subsec:light_curve} we made an attempt to separate low- and high-state $\gamma$-ray emissions of PKS 1510--089 using two-dimensional analysis with spectral index and photon flux as independent variables. Fig.~\ref{fig:2d_lowstate} shows the resulting scatter plot with mean and median points, as well as $1\sigma$, $2\sigma$, and $3\sigma$ bounding ellipses. For comparison, the final photon flux threshold from Fig.~\ref{fig:lowstate_hist} is plotted as horizontal red line.
	
It is clear that, although a slight positive correlation is present between the spectral index and the photon flux, it should not significantly affect the state separation procedure. Moreover, a strict lower bound on photon flux makes it impossible to use the bivariate normal distribution approximation (which is also clear from the displaced $n\sigma$ ellipses) and demands a more sophisticated analytical or numerical solution. We therefore conclude that the empirical PDF-based photon flux procedure described in Section \ref{subsec:light_curve} is sufficient for the low-state identification.
	
% Don't change these lines
\bsp	% typesetting comment
\label{lastpage}
\end{document}